\documentclass[aps, pre, twocolumn, showpacs,amssymb,superscriptaddress]{revtex4-1}

\usepackage{epsfig, color}
\usepackage{amsthm, amsmath, amsfonts, ae, psfrag}
\usepackage{braket}
\usepackage{bbm}
\usepackage[english]{babel}
\usepackage{epstopdf} 
\usepackage{notes2bib}
\usepackage{listings}
\usepackage{verbatim} 

\usepackage[colorlinks,
pdfpagelabels,
pdfstartview = FitH,
bookmarksopen = true,
bookmarksnumbered = true,
linkcolor = blue,
plainpages = false,
hypertexnames = false,
citecolor = blue] {hyperref}

\begin{document}

\title{Nonequilibrium phase transitions of sheared colloidal microphases: \\
	 Results from dynamical density functional theory}

\author{Daniel~Stopper}
\email{daniel.stopper@uni-tuebingen.de}
\author{Roland~Roth}

\affiliation{Institute for Theoretical Physics, University of T\"ubingen, Auf der Morgenstelle 14, 72076 T\"ubingen, Germany}

\date{\today}

\begin{abstract}

By means of classical density functional theory and its dynamical extension, we consider a colloidal fluid with spherically-symmetric competing interactions, which are well known to exhibit a rich bulk phase behavior. This includes  complex three-dimensional periodically ordered cluster phases such as lamellae, two-dimensional hexagonally packed cylinders, gyroid structures or spherical micelles. While the bulk phase behavior has been studied extensively in earlier work, in this paper we focus on such structures confined between planar repulsive walls under shear flow. For sufficiently high shear rates, we observe that microphase separation can become fully suppressed. For lower shear rates, however, we find that e.g. the gyroid structure undergoes a kinetic phase transition to a hexagonally packed cylindrical phase, which is found experimentally and theoretically in amphiphilic block copolymer systems. As such, besides the known similarities between the latter and colloidal systems regarding the equilibrium phase behavior, our work reveals further intriguing non-equilibrium relations between copolymer melts and colloidal fluids with competing interactions. 
\end{abstract}


\maketitle

\section{Introduction} \label{SecIntro}

Various recent theoretical and simulation studies have predicted that colloidal suspensions interacting via short-ranged attractive forces in addition to longer-ranged repulsive forces can self-organize into periodically ordered cluster phases (also termed in the literature inhomogeneous bulk phases, mesophases, or microphases) without any external fields \cite{Ciach2008, EdelmannRothPRE2016, PiniParola2017, Charbonneau2016_PRL, CharbonneauReview2016}. Remarkably, even for spherically-symmetric interaction potentials, the morphologies of such structures can be very complex, including lamellar (L), spherical micelles (S) on cubic lattices (e.g. BCC), hexagonally packed cylinders (HEX), or gyroids (G). These phases are formed by low- and high density domains, where within the latter local packing fractions can reach up to 40\%\,, but in low-density regimes the packing fraction can be less than 1\%\,.
Unfortunately, an experimental validation of ordered pattern formation in colloidal systems is still lacking, although the existence of unordered cluster phases and gels has been reported in experiments \cite{Stradner2004, Campbell2005, Zhang2012_SoftMatter} and computer simulations \cite{Sciortino2004_PRL_compInt, Bollinger_2016, Bomont2017, DasDhont2018}. This is probably because many long-living metastable cluster states can be present, and hence it is not clear whether periodic microphases are dynamically accessible in experiments. However, recent simulations have suggested that at least simple structures such as the lamellar phase should in principle be obtainable \cite{Charbonneau2017_JCPCom}. 

Remarkably, self-assembly of ordered phases has been observed experimentally and theoretically in systems that exhibit fundamentally different types of interactions, e.g. strongly orientation-dependent forces such as those present in block copolymers \cite{Foerster1994_copolymerClusterPhases, MatsenSchick1994_PRL}, where chain connectivity is frustrated by the immiscibility of different polymer components, or oil-water mixtures containing hydrophilic and hydrophobic ions leading to microphase-separation of the solvent \cite{TasiosDijkstra2017}. In particular, for block copolymers and colloidal suspensions intriguing quantitative similarities are observed in terms of the bulk phase behavior; more specifically, these similarities are manifested in an unique ordering of structures S $\rightarrow$ HEX $\rightarrow$ G $\rightarrow$ L $\rightarrow$ G $\rightarrow$ HEX $\rightarrow$ S upon increasing the volume fraction of the constituents. A line akin to a spinodal separates the disordered fluid-type region from microphase-separated states. This universal behavior has been rationalized by showing that both systems can be characterized by the same Landau-Brazovskii free-energy functional \cite{Ciach2013}. Hence, the formation of ordered patterns seems to be a general result of complex physical mechanisms leading to a competition between configurational energy and entropy, irrespective of its precise microscopic origin.

In experiments and simulations upon block copolymers, samples are often exposed to external forces such as confining geometries, shear flow, or electrical fields in order to pin down static, dynamic, and mechanical properties of self-assembled structures, as a comprehensive understanding of underlying microscopic mechanisms is crucial for technological and industrial applications such as drug delivery \cite{Kataoka2001, Roesler2012}, nanoscale patterning \cite{Li2006, Krishnamoorthy2006} or lithography \cite{Tang2008}. Typically, shear forces are applied in order to generate longer-ranged order of specific types of structures, as spontaneously formed configurations usually exhibit a large number of defects, or to drive phase transitions between distinct types of phases in a controlled way \cite{Register2013}. For instance, both experimental and simulation results have seen shear-induced phase transitions G $\rightarrow$ HEX \cite{Eskimergen2005, Pinna2008}.

In light of the latter and the known analogies between colloidal systems and block copolymers in equilibrium, in this work we theoretically investigate the effect of shear on (confined) colloidal microphases, as predictions from theory and simulations can play a key rule in guiding experimental investigations.
To this end, we consider a colloidal fluid exhibiting competing interactions which is confined between to planar repulsive barriers. We employ classical density functional theory (DFT) \cite{Evans1979} and its dynamical extension (DDFT) \cite{MarconiTarazona1999, ArcherEvans2004}, which form powerful and well established tools for describing both equilibrium \cite{Schmidt2000, Roth2010, Thorneywork2014JCP, reviewDFT_2016, ArcherEvans2017, GussmannRoth2017, FriesStopper2017}  and dynamic (non-equilibrium) phenomena \cite{RoyallDzubiellaSchmidtBlaaderen2007, RothRauscherArcher2009,  StopperHHGRoth2015JCP, Goddard2016, Zimmermann2016, ScacciBrader2017_MolPhhys, ScacchiArcherBrader2017}. In particular, we make use of a recently developed version of DDFT that is capable of capturing fundamental aspects of sheared colloidal suspensions such as shear-induced migration and laning transitions \cite{BraderKrueger_2011, ScacchiKruegerBrader2016, ScacchiArcherBrader2017}. 

The precise model system that we consider is a two-Yukawa fluid in addition to a hard core with diameter $\sigma$, which is a typical colloidal model system when investigating competing interactions. We employ a standard mean-field approach to describe the non-hard-core interactions \cite{HansenMcDonald2013, ArcherEvans2017}, where the hard core is modeled within Rosenfeld's fundamental measure theory (FMT) \cite{Rosenfeld1989, Roth2010}. However, note that the qualitative phase behavior is not sensitive to the precise type of the interaction -- e.g. a square-well-linear-ramp \cite{Charbonneau2016_PRL} or a patchy attraction in addition to a spherical repulsion \cite{StopperRoth2017} have also been considered.
 
In this work we focus on the L and G phases; the former structure presumably obeys the simplest morphology which minimizes numerical challenges but allows us to study fundamental aspects. The G phase, which separates space into two labyrinths of cylindrical passages that never meet, is of particular interest as such complex structures do not occur only in soft-matter systems. For instance, they are also prominent in the wings of specific types of butterflies or birds \cite{Dufresne2009, Saranathan2010} acting as biophotonic crystals responsible for structural coloring. 
The paper is structured as follows. In Sec. \ref{SecTheory} we first introduce the model system with a brief overview of the DFT formalism (Subsec. \ref{SubSec_A_Theory}). This is followed by a short discussing regarding the bulk phase behavior (Subsec. \ref{SubSec_B_Theory}), and in Subsec. \ref{SubSec_Res_MicrophasesConfinement} we put focus on the influence of confining geometries on colloidal microphases. Subsequently, in Subsec. \ref{SubSec_DDFT} we give a brief introduction to a DDFT version that is capable of describing shear flow.
In Sec. \ref{SecResults} we present our results complemented by discussions regarding the numerical implementation and the behavior of the L and G phases under shear. Finally, in Sec. \ref{SecDiscussion} we present a summary of our findings and provide an outlook for future work.

\section{Theory}\label{SecTheory}

\subsection{A DFT for the model system} \label{SubSec_A_Theory}

We consider a two Yukawa-fluid, with an attractive head close to the hard core, and a repulsive tail at longer distances. The interaction potential reads
\begin{equation}
\beta\phi(r) = \beta\phi_\text{hs}(r) + \beta\phi_\text{tail}(r)\,,
\end{equation}
where $\phi_\text{hs}(r)$ is the usual hard-sphere potential, and $\beta = 1/(k_B T)$ denotes the inverse temperature. The longer-ranged, non-hard-core interactions is ($r \geq \sigma$)
\begin{equation} \label{Eq:lr_interactions}
	\beta\phi_\text{tail}(r) = A\left(-\frac{\sigma}{r}e^{-z_1 (r/\sigma - 1)} + B\frac{\sigma}{r}e^{-z_2 (r/\sigma - 1)}\right)\,,
\end{equation}
where $A$ is a dimensionless measure of the potential depth, and $B$ fixes the amplitude ratio between the attraction and repulsion. The parameters $z_1$ and $z_2$ (with $z_1 > z_2$) control the interaction range of the attraction and repulsion, respectively.

Within the framework of DFT, the grand-potential functional of the one-body density $\rho(\mathbf{r})$ is given by \cite{Evans1979}
\begin{equation} \label{Eq:grandPotentialFunctional}
\Omega[\rho] = F_\text{id}[\rho] + F_\text{ex}[\rho] + \int \text{d}\mathbf{r}\,\rho(\mathbf{r})(V_\text{ext}(\mathbf{r}) - \mu)\,,
\end{equation}
with the chemical potential $\mu$ of the particle reservoir, and an external potential $V_\text{ext}(\mathbf{r})$. The ideal-gas intrinsic Helmholtz free-energy functional is known analytically
\begin{equation}
	\beta F_\text{id}[\rho] = \int \text{d}\mathbf{r}\,\rho(\mathbf{r})\left(\ln(\Lambda^3 \rho(\mathbf{r})) - 1\right)\,,
\end{equation}
in which $\Lambda$ is the thermal wavelength. All the particle interactions are contained in the (generally unknown) excess part $F_\text{ex}[\rho]$,
\begin{align} 
	\beta F_\text{ex}[\rho] = \beta F_\text{ex}^\text{hs}[\rho] + \frac{1}{2} \iint\text{d}\mathbf{r}'\text{d}\mathbf{r}\, \rho(\mathbf{r})\rho(\mathbf{r}')\beta\phi_\text{tail}(|\mathbf{r}-\mathbf{r}'|)\,.
\end{align}
For the hard-sphere part, for simplicity here we employ the original Rosenfeld functional \cite{Rosenfeld1989}, but obviously more accurate versions such as the White-Bear versions \cite{Roth2002, HansenGoos2006WB2} can be applied. The non-hard-core interactions are treated in a standard mean-field approach \cite{HansenMcDonald2013, ArcherEvans2017}, but with the range of the attraction extended down to the core \cite{Archer2008CompInteractions}. 
The equilibrium density profile $\rho_\text{eq}(\mathbf{r})$ can then formally be obtained from minimizing $\Omega[\rho]$ w.r.t. the density, which yields the formal expression
\begin{equation} \label{Eq_rhoEq}
	\rho_\text{eq}(\mathbf{r}) = \rho_b\exp(-\beta V_\text{ext}(\mathbf{r}) + c^{(1)}(\mathbf{r}) + \beta\mu_\text{ex})\,,
\end{equation}
where $\mu_\text{ex}$ denotes the excess part of the chemical potential $\mu$, and $\rho_b = N/V$ the reservoir bulk density. The corresponding reservoir packing fraction is $\eta = \rho_b \pi \sigma^3 / 6$. \\

However, Eq. \eqref{Eq_rhoEq} has to be solved numerically, e.g. using a Picard iteration scheme, but more sophisticated versions can be employed \cite{EdelmannRothJCP2016}. The quantity $c^{(1)}(\mathbf{r})$ is referred to as the one-body direct correlation function, defined via
\begin{equation}
	c^{(1)}(\mathbf{r}) = - \beta\frac{\delta F_\text{ex}[\rho]}{\delta\rho(\mathbf{r})}\,,
\end{equation}
which can be viewed as a (negative) local chemical potential.
The system is influenced by two planar repulsive walls described by the following external potential ($H_2 > H_1$)
\begin{equation} \label{Eq_DefVext}
V_\text{ext}(z) = V_0
\begin{cases}
1~;~~\text{if}~ z < H_1~\text{or}~z>H_2\,, \\
e^{-(z - H_1)^2/(4\sigma^2)} + e^{-(z-H_2)^2/(4\sigma^2)}~;~\text{else}\,,
\end{cases}
\end{equation}
with $V_0 = 50 k_B T$. The quantity $H \equiv H_2 - H_1$ controls the distance between the walls.

\begin{figure}[t!] 
	\centering
	\includegraphics[width = 0.5\textwidth]{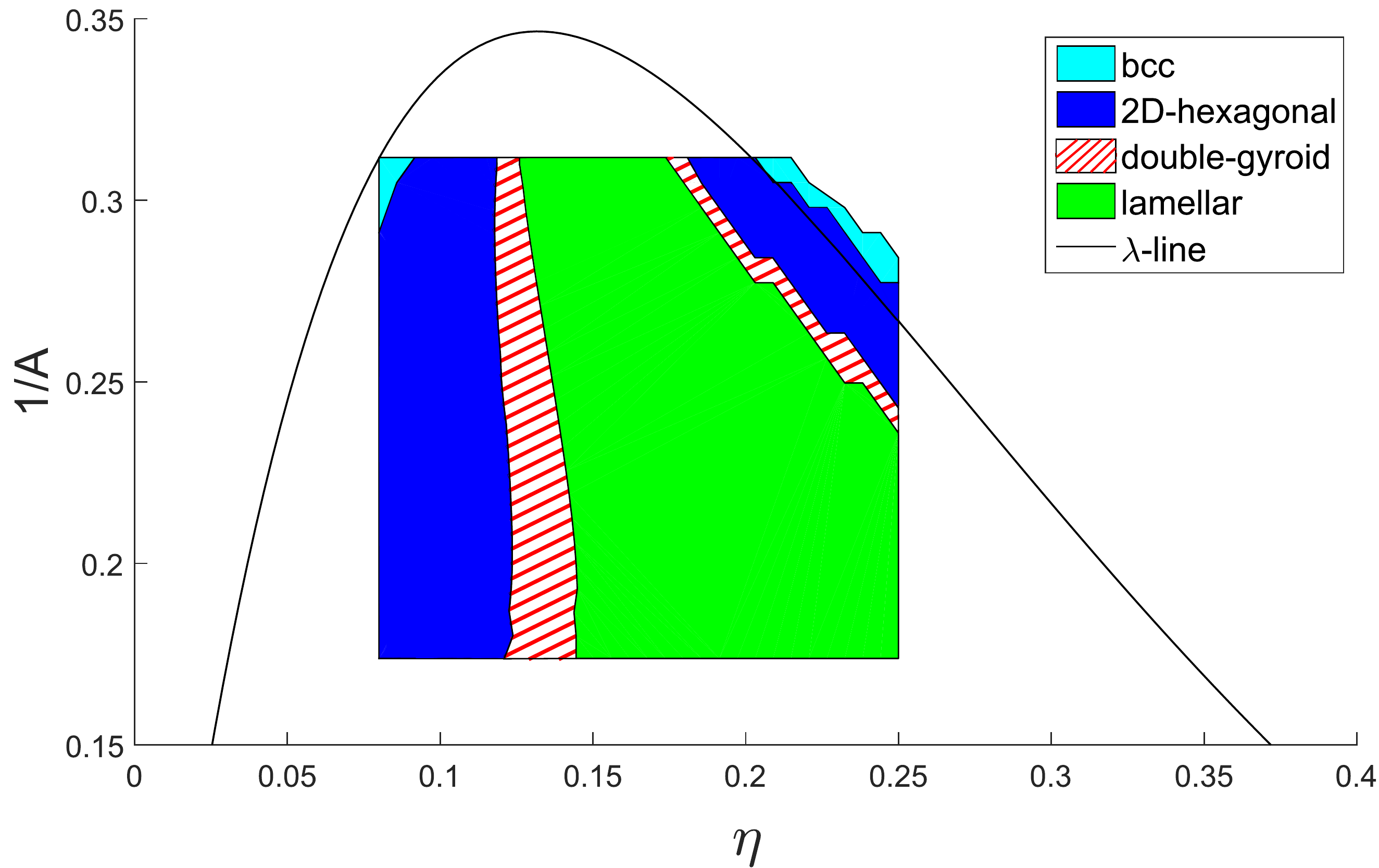} 
	\caption{ Bulk phase diagram as borne out by DFT for the present model system, reproduced from Ref. \cite{EdelmannRothPRE2016}. $\eta = \pi \rho_b \sigma^3 / 6$ denotes the reservoir packing fraction, and $1/A$ is the inverse interaction strength, which can be viewed as an effective temperature (cf. Eq. \eqref{Eq:lr_interactions}). \label{Fig_phaseDiag}}
\end{figure} 

\subsection{Bulk phase behavior} \label{SubSec_B_Theory}

The typical bulk phase behavior of colloidal systems with competing interactions has been studied in detail in various previous studies employing Landau-type theories \cite{Ciach2008, Ciach2013}, density functionals \cite{Archer2007CompInteractions, Archer2008CompInteractions, EdelmannRothPRE2016, PiniParola2017}, or computer simulations \cite{Charbonneau2016_PRL, CharbonneauReview2016}. Therefore, we do provide only a compact repetition of known results so far. 

For sufficiently chosen parameters $z_1$, $z_2$, and $B$ in the interaction potential Eq. \eqref{Eq:lr_interactions} introduced in Sec. \ref{SubSec_A_Theory}, one finds a region in the phase diagram (e.g. in the temperature-density plane, where $1/A$ takes the role of an temperature) preempting the usual gas-liquid binodal in which the homogeneous bulk state $\rho(\mathbf{r}) = \rho_b$ is unstable w.r.t. arbitrary small density fluctuations. This instability region is enclosed by the so-called $\lambda$ line. Within mean-field theory, the $\lambda$ line is related to a diverging peak in the static structure factor $S(k)$  of the homogeneous system for a small but non-zero wavenumber $0 < k_c \ll 2\pi/\sigma$ -- recall that the gas-liquid-spinodal, which describes the region in which a macroscopic phase separation into a gas and a liquid inevitably occurs is related to a divergence in $S(k)$ at $k=0 $. 
The static structure factor analytically can be calculated via \cite{HansenMcDonald2013}
\begin{equation}
S(k) = \frac{1}{1-\rho_b\widehat{c}(k)}\,,
\end{equation} 
where $\widehat{c}(k)$ for the present system is the three-dimensional (3D) Fourier-transform of the direct pair-correlation function $c(r)$ given by
\begin{equation}
c(r = |\mathbf{r} - \mathbf{r}'|) = -\left.\beta\frac{\delta^2 F_\text{ex}[\rho]}{\delta \rho(\mathbf{r})\delta \rho(\mathbf{r}')}\right|_{\rho = \text{const.}}\,.
\end{equation}

Within the region enclosed by the $\lambda$ line, the present mean-field DFT predicts that a periodically ordered inhomogeneous cluster state can lower the free energy of the system compared to the homogeneous bulk. A more detailed analysis reveals that in specific regions lamellar, cylindric-hexagonal tubes, gyroid, or spherical-micelles-like structures are thermodynamically the most stable phases thereby showing a unique ordering, see Fig. \ref{Fig_phaseDiag}. Importantly, this pattern-formation is also predicted by computer simulations \cite{Charbonneau2016_PRL}. 
Note also that, in addition to the $\lambda$ line where $S(k_c)$ diverges, there exists a typically broader region in the phase diagram, where $S(k_c)$ exhibits a growing, but finite peak at low wave numbers. This may be related to unordered cluster phases provided that the peak height and width exceeds a certain threshold \cite{Sciortino2004_PRL_compInt, Bollinger_2016, Bomont2017, DasDhont2018}.

\begin{figure*}[t!] 
	\centering
	\includegraphics[width = 15cm]{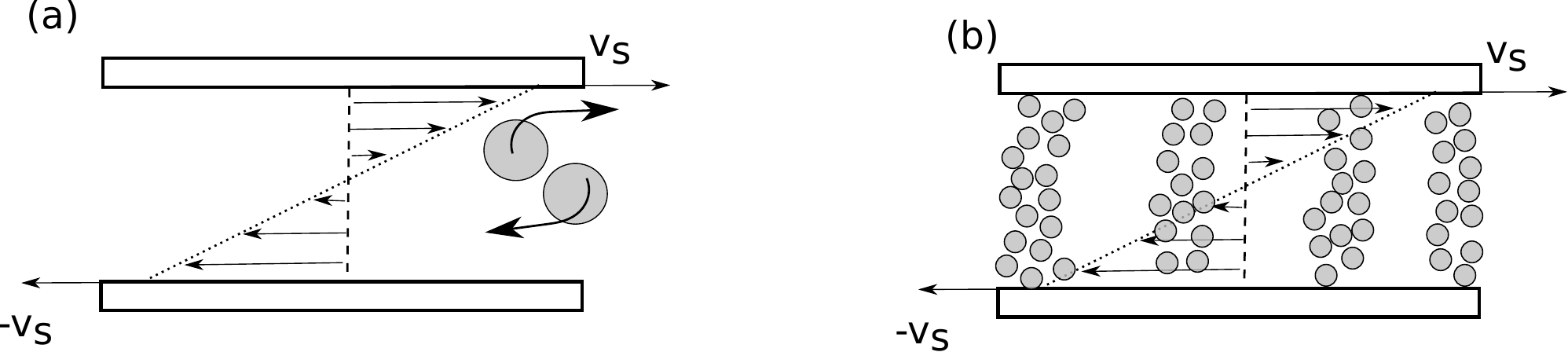} 
	\caption{(a) Microscopic picture of the effect of shear flow on colloidal suspensions: Particles that are subject to distinct shear velocities tend to overtake each other, which generates a distortion of two-body correlation functions.  (b) A schematic picture of a periodically ordered cluster state under shear where two effects take place: in addition to the microscopic \lq overtake-mechanism\rq\, shown in (a), the high-density domains themselves are getting deformed due to a non-vanishing density gradient along the flow direction.}
	\label{Fig:shear_microscopics}
\end{figure*}

\subsection{Colloidal microphases in confined geometries} \label{SubSec_Res_MicrophasesConfinement}

Facing the rich bulk phase behavior of the present system, a confining geometry introduces additional parameters, thereby making the problem considerably richer and more complex. For two planar repulsive barriers, we expect at least two parameters to become important: First, the separation $H$ of the walls will influence the phase behavior of the system -- e.g., if $H < L_0$, where $L_0$ denotes the typical bulk length scale (i.e., periodicity length) of a certain cluster morphology, it is possible to suppress the formation of this specific structure. In particular, if the separation of the walls becomes sufficiently small such that it is no longer commensurate with the periodicity length of any possible cluster states, it is possible to completely suppress microphase separation. This, for instance, has been demonstrated in a previous work addressing two-dimensional systems with competing interactions confined between planar walls \cite{Archer2008PRE_2dCompInt}.
A second quantity that becomes worth studying is the preferred orientation $\mathcal{P}$ of structures relative to the walls. For instance, it has been shown experimentally that in copolymer systems the orientation of G phases can be controlled by tuning substrate-polymer interactions \cite{Park2017}. 
Thus, taking all parameters into account, one finds a four-dimensional phase space spanned by the quantities ($A$, $\eta$, $H$, $\mathcal{P}$) if the parameters $z_1$, $z_2$, and $B$ in Eq. \eqref{Eq:lr_interactions} are kept constant. In principle, it is then possible to deduce the most thermodynamic stable cluster phase at each point in phase space, by comparing the grand potentials of all structures -- although, from a practical point of view, this seems to be out of reach within reasonable computational effort as a full three-dimensional minimization of DFT (even for one point in phase space) is numerically intensive \cite{EdelmannRothJCP2016,EdelmannRothPRE2016, PiniParola2017}.
In this work we therefore do not focus on determining the full phase behavior; in fact, we pick up a state point within the bulk phase diagram (e.g., where the gyroid is predicted to be stable) and choose the external potential such that the wall separation is $H = n L_0$ where $n \in \mathbb{N}$ (typically $n = 2,3,4$). The orientation $\mathcal{P}$ is chosen such that two planes of the unit cell are faced parallel to the wall barriers -- irrespective of whether another orientation may be energetically favored.

\subsection{Dynamic DFT and shear flow} \label{SubSec_DDFT}

For particles that are subject to overdamped Brownian motion, dynamic DFT (or DDFT) \cite{MarconiTarazona1999, ArcherEvans2004} forms a valuable theoretical tool for investigating dynamic (non-equilibrium) phenomena, as was demonstrated in various studies comparing DDFT to experiments and/or computer simulations, see, e.g., Refs. \cite{RoyallDzubiellaSchmidtBlaaderen2007,  StopperHHGRoth2015JCP, Zimmermann2016, StopperThorneywork2018}. The theory assumes that correlations out of equilibrium instantaneously (i.e., at each point in time) can be mapped to a corresponding equilibrium situation described by a suitable external potential.
In particular, standard DDFT has been extended to describe flowing solvents around external potentials \cite{RauscherKrueger2007}  (e.g. a colloid immersed in a sea of non-interacting advected polymers). The key equation of the advected-DDFT reads
\begin{equation} \label{EqDDFT_flowfield}
\frac{\partial \rho(\mathbf{r},t)}{\partial t} + \nabla\cdot(\mathbf{v}(\mathbf{r})\rho(\mathbf{r},t)) = D_0\nabla\left(\rho(\mathbf{r},t)\nabla \frac{\delta \beta \Omega[\rho]}{\delta \rho(\mathbf{r}, t)}\right)\,,
\end{equation}
where $\mathbf{v}(\mathbf{r})$ is the flow field, $D_0 = \sigma^2/\tau_B$ is the Stokes-Einstein diffusion coefficient, and $\tau_B$ is the Brownian time. $\Omega[\rho]$ is precisely the \textit{equilibrium} grand-potential functional \eqref{Eq:grandPotentialFunctional}. \\
For fluids with \lq simple\rq\, types of interactions, the particle density confined between planar walls depends only on the direction normal to these walls. For the present external potential given by Eq. \eqref{Eq_DefVext} this means $\rho(\mathbf{r}) = \rho(z)$. Now assume a simple shear profile of the solvent, with flow direction parallel to the barrier boundaries, such that the particle motion is not directly hindered by the external potential. A suitable choice in light of Eq. \eqref{Eq_DefVext} is 
\begin{equation} \label{Eq_Def_shearFlow}
\mathbf{v}(\mathbf{r}) = \dot{\gamma} (z - H/2) \hat{e}_x = \text{Pe}\frac{D_0}{R^2}(z - H/2) \hat{e}_x\,,
\end{equation}
where $\dot{\gamma}$ is the shear rate, and $\text{Pe} = \dot{\gamma} R^2/D_0$ denotes the Peclet number which we use as a synonym to shear rate, as the particle hard-core radius $R$ is kept constant in this work. Unfortunately, for such flow profiles, Eq. \eqref{EqDDFT_flowfield} gives rise to $\nabla\cdot(\mathbf{v}(\mathbf{r})\rho(\mathbf{r}, t)) = 0$, which immediately results in the key equation of standard DDFT
\begin{equation}
\frac{\partial \rho(\mathbf{r},t)}{\partial t} = D_0\nabla\left(\rho(\mathbf{r},t)\nabla \frac{\delta \beta \Omega[\rho]}{\delta \rho(\mathbf{r}, t)}\right)\,.
\end{equation}

However, for particles with a repulsive (hard) core this cannot be correct (see Fig. \ref{Fig:shear_microscopics} (a))  \cite{BraderKrueger_2011}: particles that are exposed to distinct solvent velocities tend to overtake each other, resulting in a pressure component normal to the walls. In experiments it has been demonstrated that this effect can lead to shear-induced migration at walls, and, when shear becomes sufficiently strong, particles can self-organize into characteristic layers reminiscent of a crystal (so-called laning transition) in order to slide past one another more efficiently \cite{AckersonPusey1988, Besseling2012}.
This failure of DDFT has been traced back to the key assumption of the latter, namely that equilibrium sum rules hold also out of equilibrium \cite{BraderKrueger_2011}. This is equivalent to neglecting any coupling between particle interactions and external flow fields. A method that bypasses this shortcoming was first suggested in \cite{BraderKrueger_2011}, and a very promising correction to DDFT recently was derived, which is exact in the dilute limit and in principle is applicable to all kind of particle interactions \cite{ScacchiKruegerBrader2016, ScacchiArcherBrader2017}. Here, the flow field $\mathbf{v}(\mathbf{r})$ in the left-hand side of Eq. \eqref{EqDDFT_flowfield} is corrected according to $	\mathbf{v}(\mathbf{r})\rightarrow\mathbf{v}(\mathbf{r})+\mathbf{v}_\text{fl}(\mathbf{r})$, where 
\begin{equation} \label{EqDef_vflow}
\mathbf{v}_\text{fl}(\mathbf{r}) = \int \text{d}\mathbf{r}'\,\rho(\mathbf{r}') \mathbf{K}(\mathbf{r}-\mathbf{r}')\,,
\end{equation}
and $\mathbf{K}(\mathbf{r})$ is the so-called flow kernel, which implements the physical mechanism of one particle moving past another. 

For the present system, however, we have a different situation as the density of a modulated phase may depend on all three space variables $x,y,z$. This can result in a non-zero advection term $\nabla\cdot(\mathbf{v}(\mathbf{r})\rho(\mathbf{r},t))$ due to non-vanishing density gradients \textit{along} the flow direction as is depicted schematically in Fig. \ref{Fig:shear_microscopics} (b). Thus, we have a superposition of two effects: the flow drives a deformation of the structures themselves, and, within high-density domains of the latter, particles undergo the microscopic \lq overtake\rq\, mechanism discussed previously. Thus, a key-task is to determine the flow kernel $\mathbf{K}$ -- 
in this paper, as a first-order approximation, we employ the flow kernel for hard spheres \cite{ScacchiKruegerBrader2016}, as it has an analytic form that can be Fourier-transformed also analytically (see Appendix \ref{Appendix_B}). This means that Eq. \eqref{EqDef_vflow} can be solved very efficiently using Fourier-methods during the numerical time integrations of Eq. \eqref{EqDDFT_flowfield}, which is a crucial aspect in three dimensions. Moreover, as we will demonstrate in the next Section \ref{SecResults}, the flow kernel becomes negligible in the limit of low shear rates Pe < 1.

As a side remark we finally note that in this work we solve the DDFT in three dimensions up to long times $t \sim 10^2 \tau_B$, which requires an adequate solver for elliptic partial differential equations that is accurate on the one hand, but also allows for sufficiently large time steps $\Delta t$ in order to obtain reasonable computation times. To this end, we employ a method called exponential time differencing \cite{CoxMatthews2002},  which during our investigations turned out to be very accurate for diffusive equations. The formalism is presented in Appendix \ref{Appendix_A}. 
	We tested the method by integrating several equilibrium density profiles (e.g., the G structure) forward in time without any shear up to $t\sim 10^3 \tau_B$, employing (i) an explicit Euler-forward integrator, (ii) an implicit Crank-Nicholson integrator, and (iii) the exponential time differencing method. For all three methods we used a time step of $\Delta t = 10^{-4} \tau_B$. In equilibrium without any shear, we expect that $\partial \rho(\mathbf{r},t)/\partial t = 0$ irrespective of up to which times the DDFT equations are numerically solved for -- however, we found that only method (iii) did not give rise to a systematic decay of structure in the density profiles after very long integration times. In particular, the common but simple Euler-forward algorithm turned out to be highly unstable with the chosen time step, manifested in numerical divergences after relatively short times after shear set in. It could only be stabilized by choosing a very small time step of $\Delta t = 10^{-6} \tau_B$.   \\

\section{Results and discussion} \label{SecResults}

\subsection{Lamellar and gyroid structures confined by repulsive walls} \label{SubSec_GyroidWall}

\begin{figure}[t!] 
	\centering
	\includegraphics[width = 8cm]{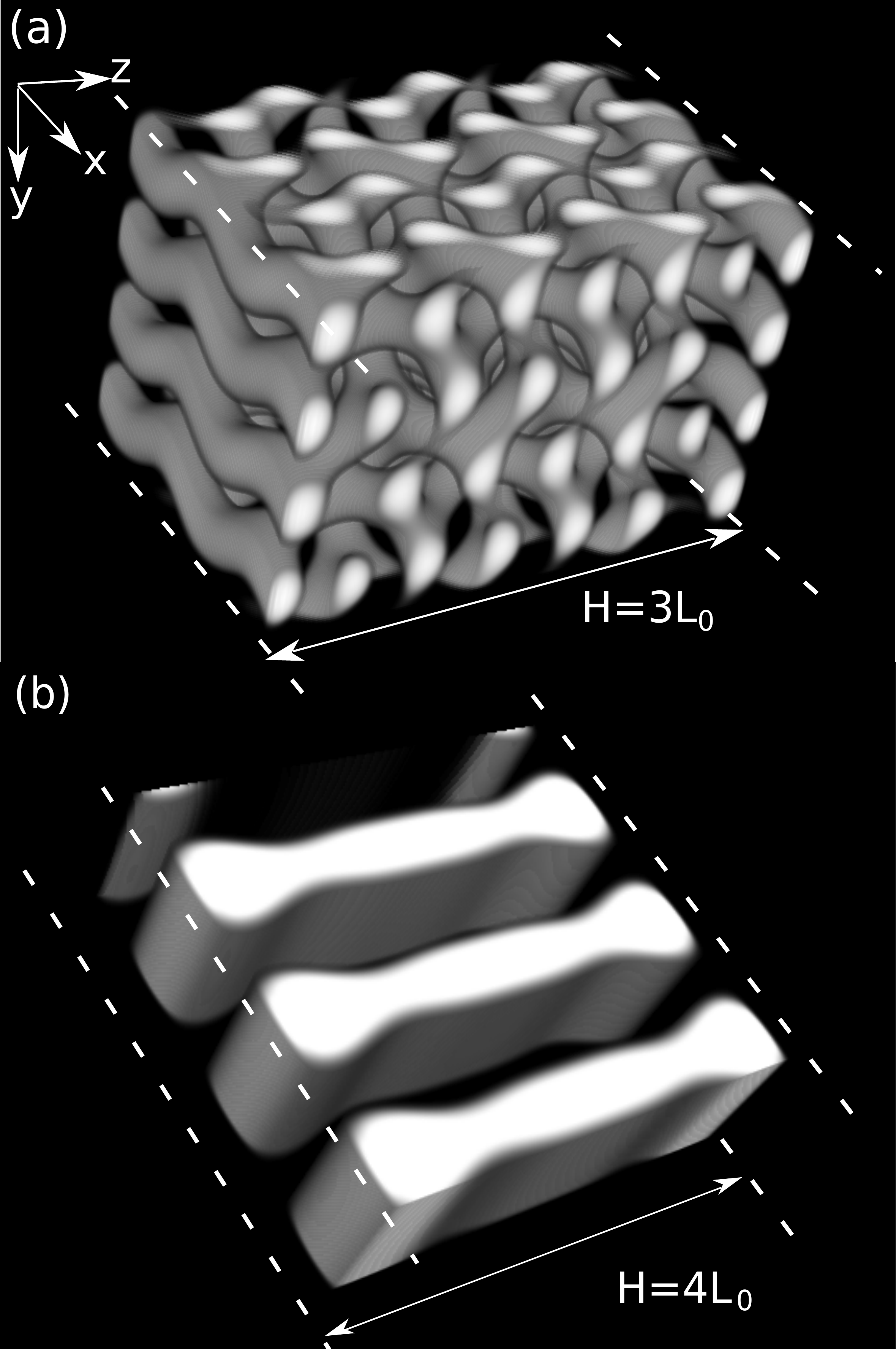} 
	\caption{(a) Gyroid structure between repulsive walls with distance $H = 3L_0$ ($z$-axis is perpendicular to the walls, dashed lines are guides to the eye indicating the latter) as obtained from numerically solving Eq. \eqref{Eq_rhoEq} (see text for further details). Periodic boundary conditions are applied along $x$- and $y$-directions; white means highest particle density, and black zero density. (b) Lamellar-type structure with planes perpendicular to the walls with separation $H = 4 L_0$. The coordinate system applies to (a) and (b). The reservoir packing fraction is $\eta = 0.12$ in (a), and $0.15$ in (b). The inverse attraction strength is $1/A = 0.22$ in both cases.}
	\label{Fig:gyroid_wall}
\end{figure}

In Fig. \ref{Fig:gyroid_wall} (a) we show a gyroid structure confined between repulsive walls according to Eq. \eqref{Eq_DefVext} with distance $H = 3L_0$ and $L_0 \approx 25 \sigma$, and in (b) a lamellar phase with planes oriented perpendicular to the walls (this type of orientation we will refer to as \lq transversal\rq\,), where $H = 4L_0$ and $L_0 \approx 8 \sigma$. There is no shear force applied to the system at this point. Note also that for the present structures the periodicity length $L_0$ is the same in all three spatial directions. The coordinate system holds for (a) and (b)  and furthermore for all subsequent figures that show three-dimensional data. The dashed white lines are guides to the eye indicating the position of the repulsive walls. White areas correspond to highest particle densities ($\rho(\mathbf{r})\sigma^3 \approx 0.8$) and black means zero density. Periodic boundary conditions are applied in $x$- and $y$-directions. The reservoir bulk packing fraction is $\eta = 0.12$ for the gyroid, and $\eta = 0.15$ for the lamellar phase. In both cases and throughout this work, we fix $z_1 = 1.0$, $z_2 = 0.5, B = 0.2$, and the inverse attraction is $1/A = 0.22$. 
We see that the gyroid forms closed tubes close to the repulsive barriers, but its morphology is not distorted at the center between the repulsive barriers; the perpendicular lamellae form characteristic bulges at the walls and its density distribution is invariant along the $y$-direction. Interestingly, while we find the rather complex G and the more simple parallel-oriented L phase also for small wall separations with $H \sim L_0$ (although here the G structure is significantly affected by the walls), the transversal L phase, as well as the spherical BCC and HEX phases, could not be stabilized for $H \lesssim 4 L_0$. \\

The structures are obtained by numerically solving Eq. \eqref{Eq_rhoEq} massively in-parallel on graphics cards via a standard Picard iteration scheme in presence of the walls. It is convenient to perform three-dimensional DFT calculations on computing devices that allow for a high parallelization, as standard sequential or slightly parallelized algorithms lead to unreasonably high effort in terms of computing time  \cite{StopperRoth2017JCP}. 
 One typically has to employ an initial guess $\rho_\text{init}(\mathbf{r})$ for the density that already resembles the desired microphase \cite{EdelmannRothPRE2016}. This is due to the fact that the free-energy landscape can be very complex, i.e., within the region enclosed by the $\lambda$ line there can exist a lot of metastable cluster states. This means that during the Picard iteration the system easily can become trapped in a local minimum, which may be far away from the desired structure when simply employing a random noise term above the bulk profile as an initial input. For instance, a suitable guess for the transversal L structure reads $\rho_\text{init}(\mathbf{r}) = \rho_b(1 +  \gamma\sin(2\pi x/ L_0))\exp(-\beta V_\text{ext}(z))$, where $\gamma$ controls the strength of the perturbation and typically $\gamma = 0.1$ is sufficient. For the G, a more complex approximation based on Fourier-expansions has to be used; it can be found in Ref. \cite{vonSchnering1991}. The optimal periodicity length $L_0$ of a specific microphase can be found by minimizing the grand potential functional $\Omega[\rho]$ in Eq. \eqref{Eq:grandPotentialFunctional} w.r.t. both the density $\rho(\mathbf{r})$ and the size of the unit cell without any external potential \cite{EdelmannRothPRE2016,StopperRoth2017}.\\
The displayed G structure has a total volume of $V = L_x \times L_y \times L_z =  2 L_0 \times 2 L_0 \times 3 L_0 = 12 L_0^3$. In Ref. \cite{StopperRoth2017JCP} we have shown that typically 10 points per hard-core radius $R$ are necessary in order to guarantee that packing effects on the one-particle level are properly described, provided that particle densities are sufficiently low such that the system is still in a non-crystal phase. However, for the gyroid shown in Fig. \ref{Fig:gyroid_wall} (a) a resolution of $\sim$ 10 points/$R$ would require at least a total amount of $1024\times1024\times1536$ grid points ($L_0 \approx 50 R$). The (dynamic) DFT calculations shown in this work have been performed on high-performance graphics cards with an memory amount of 12 Gigabytes, which allow for domains consisting of maximal $256\times 256 \times 512$ grid points. This corresponds to a spatial resolution of 3 points$/R$. Thus, in order to describe a domain consisting of $1024\times1024\times1536$ grid points, an amount of memory of roughly $500$ Gigabytes is needed. However, up to this date, such large computing resources are not yet available on graphics cards. 
We have verified that the overall G structure essentially is unaffected by the coarse grid resolution, but single-particle packing effects that may occur within high-density domains cannot be resolved anymore. This we concluded by comparing the bulk gyroid structure of a large system with $V = (3 L_0)^3$ to a smaller system with $V = L_0^3$, where for the latter situation a resolution of 10 points$/R$ was employed (with $512^3$ grid points in both cases; this was possible since a purely equilibrium DFT-program is less memory intensive than a DDFT-program).

In contrast, the periodicity length of the lamellar phase ($L_0 \sim 8 \sigma$) is much smaller compared to the G phase, hence allowing for a spatial resolution of $\sim$ 10 points per radius. More precisely, for the situation shown in Fig. \ref{Fig:gyroid_wall} (b) the total volume is $V = 3 L_0 \times L_0 \times 4 L_0$ with a domain size of $384\times128\times512$ grid points. 

\subsection{Lamellar phases under steady shear} \label{SubSec_Res_LamellarShear}

\subsubsection{Parallel orientation and effects of high shear rates} \label{SubSubSec_parallel_L}

\begin{figure}[t!] 
	\centering
	\includegraphics[width = 8.5cm]{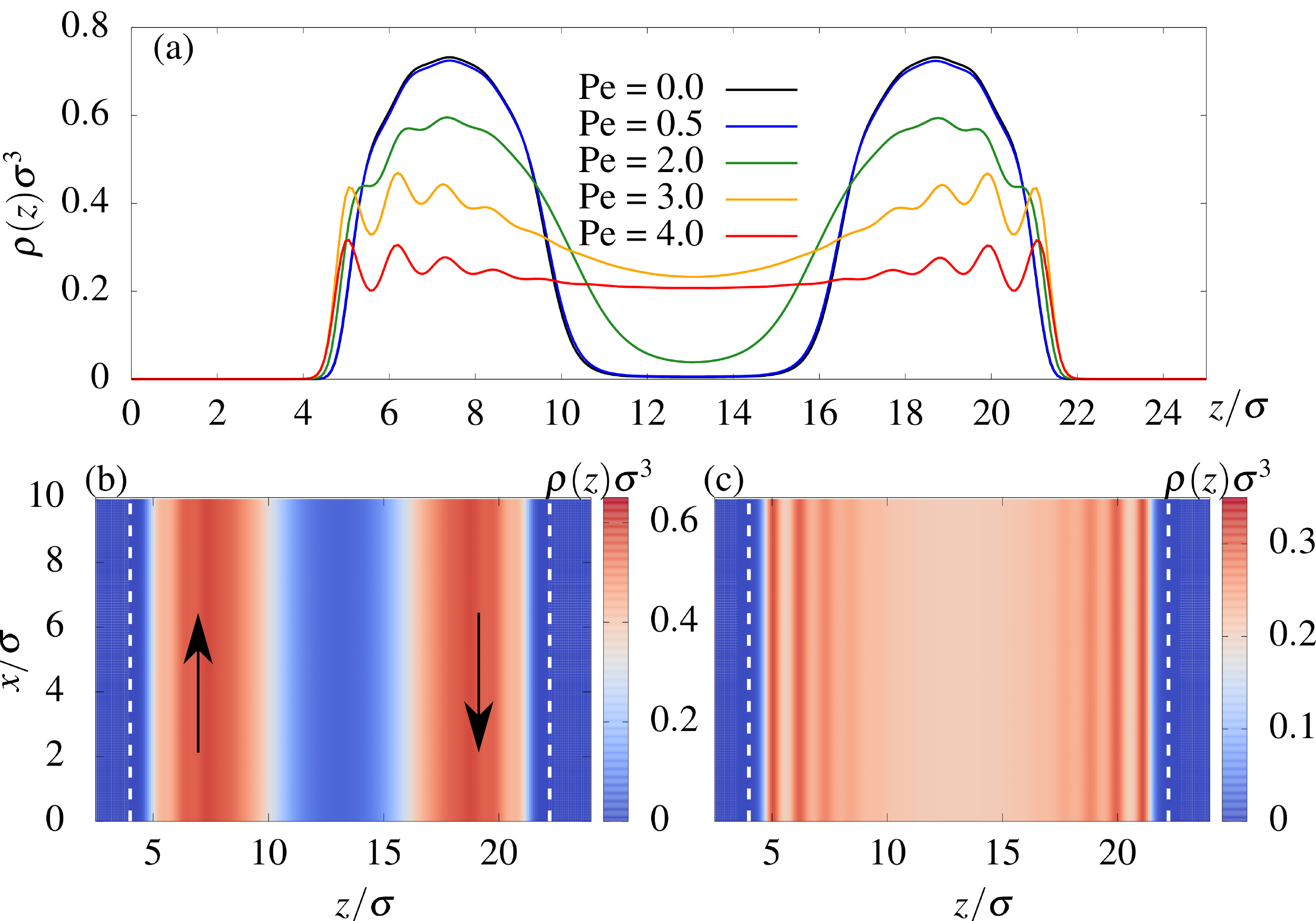} 
	\caption{(a) Steady-state density profiles $\rho(z)\sigma^3$ along the $z$-axis for several Peclet numbers: Pe = 0.5 (blue), 2.0 (green), 3.0 (yellow), and 4.0 (red). The black line shows the respective equilibrium density profile without any shear. Lower figures plot the corresponding two-dimensional cut of the 3D density in the $x$-$z$-plane for Pe = 2.0 (b) and Pe = 4.0 (c). Blue means zero, and red highest density. The dashed lines in (b) and (c) are guides to the eye marking the locations of the repulsive barriers, and the black arrows indicate the shear flow.}
	\label{Fig:lamellar_highPe}
\end{figure}

We start with considering the L phase, with the lamellae being oriented parallel to the repulsive walls. As for the transversal configuration, the reservoir packing fraction is set to $\eta = 0.15$. For the given potential parameters (cf. Sec. \ref{SubSec_GyroidWall}) this structure corresponds to the most stable thermodynamic state, whereas the transversal lamellae shown in Fig. \ref{Fig:gyroid_wall} (b) are metastable w.r.t. the former (though they represent also a local energetic minimum of the system). This can be concluded by comparing the grand potentials of these two structures. The separation of the walls is kept constant at $H = 2 L_0$. Starting with an equilibrated density profile, we integrated $\rho(\mathbf{r})$ forward in time according to Eqs. \eqref{EqDDFT_flowfield} -- \eqref{EqDef_vflow} employing a time step of $\Delta t = 10^{-4} \tau_B$. \\

\begin{figure}[t!] 
	\centering
	\includegraphics[width = 8cm]{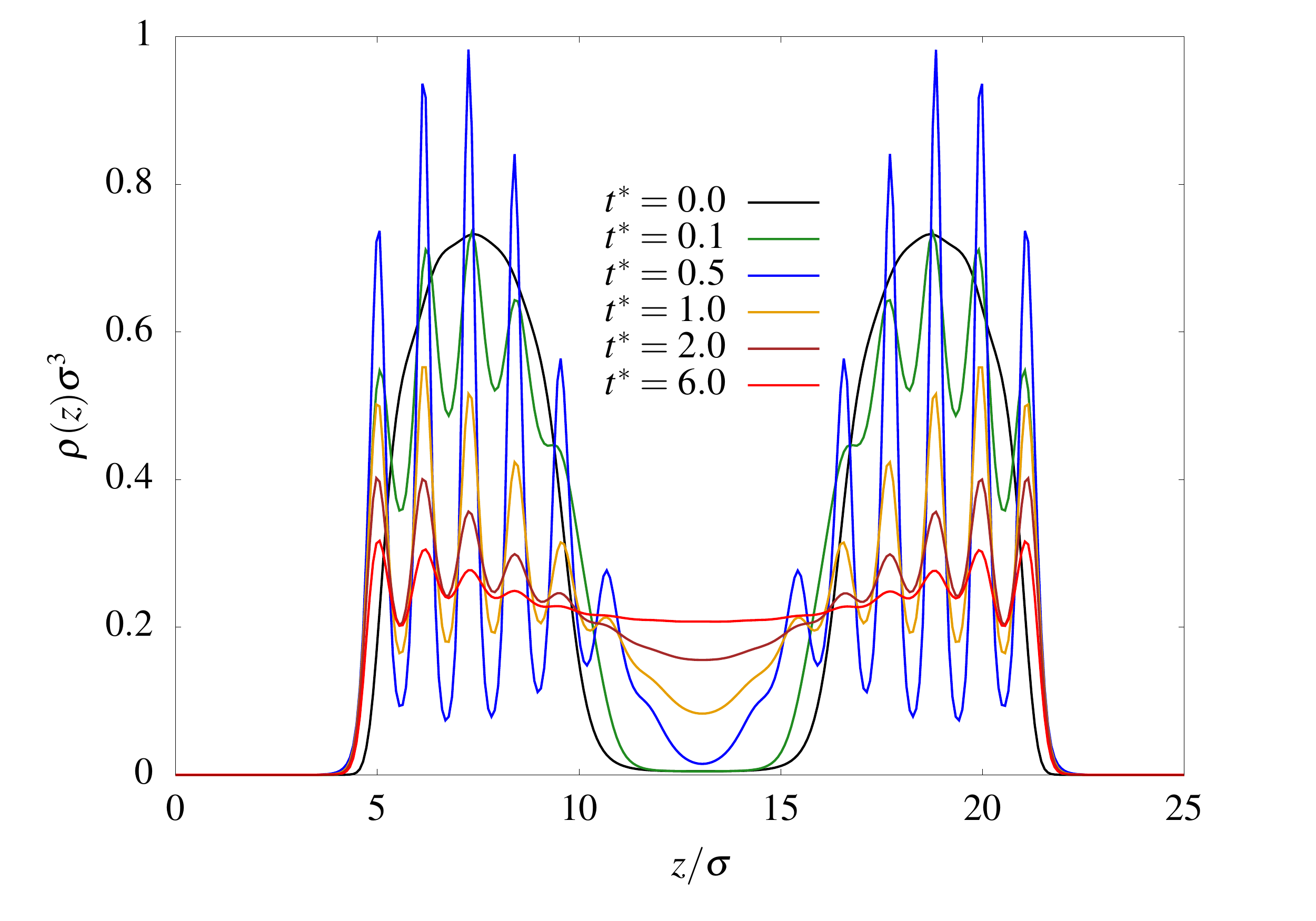} 
	\caption{Kinetic pathway of the transition from a parallel lamellar- to the dissolved phase at Pe = 4.0 for times $t^* = 0.0$ (black), 0.1 (green), $0.5$ (blue), 1.0 (orange), 2.0 (brown), and 6.0 (red). The latter is already nearly indistinguishable from the steady-state profile shown in Fig. \ref{Fig:lamellar_highPe} (a). }
	\label{Fig:lamellar_kinetics}
\end{figure}

The parallel oriented lamellar phase is most suitable to pin down effects of the flow kernel $\mathbf{K}(\mathbf{r})$, as the density varies only normal to the walls, i.e. $\rho(\mathbf{r}) = \rho(z)$. Hence, when the flow kernel is neglected, a simple shear flow has no impact on the density as the advection term in Eq. \eqref{EqDDFT_flowfield} vanishes identically. In Fig. \ref{Fig:lamellar_highPe} (a) the resulting steady-state profiles are shown along the $z$-axis for four different shear rates Pe = 0.5 (blue), 2.0 (green), 3.0 (yellow) and 4.0 (red). For the lowest shear rate, we see that the flow kernel only has a minimal impact on the density profile --it is nearly indistinguishable from the equilibrium state without shear (black line). This is to be expected, since for shear rates Pe $< 1$, the single contributions to the flow kernel become very small (all but one scale with $\sim\text{Pe}^2$). Note that although here we only considered the flow kernel for hard spheres, this argument does not change even when having access to a flow kernel that fully treats the longer-ranged interactions. Thus, for sufficiently small shear rates, the flow kernel may be omitted. This is particularly important when considering density distributions that give rise to a non-vanishing contribution in the advection term $\nabla \cdot (\mathbf{v}(\mathbf{r}, t)\rho(\mathbf{r},t))$, as we will see in the subsequent subsection \ref{SubSec_transversalLamellar}. 
However, with increasing Pe, microphase separation becomes increasingly suppressed, and particles in the vicinity of the wall boundaries become more localized. This is in line with shear-induced migration of the particles. In Figs. \ref{Fig:lamellar_highPe} (b) and (c) the respective density distribution is plotted in the $x$-$z$-plane for Pe = 2.0 (b) and 4.0 (c), where blue means zero and red corresponds to high-density domains. Figure $\ref{Fig:lamellar_kinetics}$ displays the kinetic pathway of the L phase driven out of equilibrium with Pe = 4.0. For short times $t^* = t/\tau_B$ after shear sets in, the high density domains exhibit sharply peaked one-particle correlation peaks, most pronounced for $t^* = 0.5$ (blue curve). As time goes by, the region in between the lamellae is continuously filled up, indicating that clusters are dissolved due to shear. At the same time, the correlation peaks become less pronounced, which finally yields the steady-state profile at $t^* \approx 8 $. See \cite{SupplementalMaterial1}  for a movie illustration in which the behavior of the parallel-oriented L phase under different shear rates up to $t^* = 8.0$ (color-code corresponds to the steady-state profiles shown in Fig. \ref{Fig:lamellar_highPe} (a)) is demonstrated.

This behavior can be understood as follows. Due to the linear shear profile, particles located close to the walls are moving faster than particles located at the center. This generates a pressure normal to the walls, and as a result, for sufficiently high shear, cluster formation is completely suppressed. In a recent work Scacchi \textit{et al.} derived stability thresholds for the laning transition as a function of particle interaction strength and bulk density \cite{ScacchiArcherBrader2017}. The strong localizations that are apparent in Fig. \ref{Fig:lamellar_kinetics} in the high-density domains for short times indicate the onset of such a laning transition -- for longer times, when regions with high density become increasingly flattened out, one continuously moves out of the instability region in which laning occurs. When shear is switched off after reaching the steady-state profile, the particles self assemble back into a (lamellar) cluster phase. 

In the following subsections investigating the behavior of a transversal oriented L phase (Sec. \ref{SubSec_transversalLamellar}) and G phase (Sec. \ref{SubSec_gyroid}) under shear, we will consider only shear rates with Pe < 1. For these more complex cluster morphologies it turned out that the numerics for Pe > 1  can become challenging, manifested in numerical instabilities and divergences; the one-particle peaks that emerge upon high shear can become even more localized than for the simple parallel L phase shown in Fig. \ref{Fig:lamellar_kinetics}. In order to bypass these difficulties, presumably (i) additional tensorial corrections to the original Rosenfeld functional should be taken into account, since it is well known to yield divergences when the particle density becomes too localized \cite{Rosenfeld1997, Tarazona2000}, and (ii) a higher grid resolution beyond the maximal 10 points per hard-core radius $R$ employed herein may be necessary. Unfortunately, these two points could not be addressed in this work due to limitations in terms of the available amount of graphics memory.

\begin{figure}[t!] 
	\centering
	\includegraphics[width = 8.5cm]{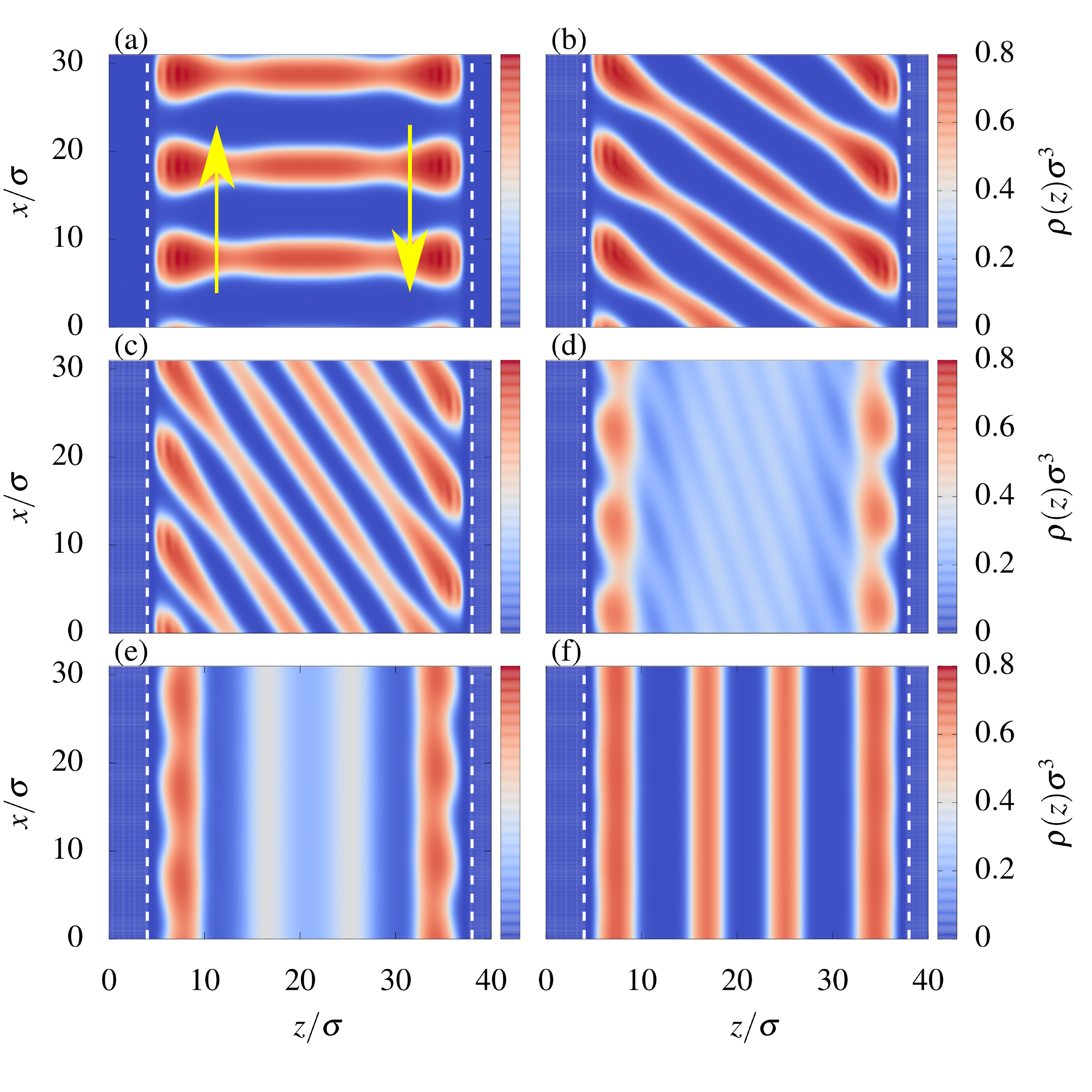} 
	\caption{Transition of an initially transversal to a parallel L phase under shear flow with Pe = 0.25 for times $t^* = 0.0$ (a), 0.50 (b), 1.0 (c), 2.0 (d), 4.0 (e) and 8.0 (f). The dashed lines are guides to the eye marking the position of the repulsive walls, and yellow arrows in (a) indicate the shear flow. In \cite{SupplementalMaterial1} a movie illustration can be found.}
	\label{Fig:lamellar_perp_kinetics}
\end{figure} 
 
\subsubsection{Transversal orientation} \label{SubSec_transversalLamellar}

As a second case, we consider a lamellar phase where planes of the lamellae are located normal to the wall, and where the flow direction is also normal to the lamellae (cf. Fig. \ref{Fig:gyroid_wall} (b)). The density distribution has a non-vanishing gradient \textit{along} the flow direction ($x$-direction), thus the advective term in Eq. \eqref{EqDDFT_flowfield} gives a non-zero contribution. In Fig. \ref{Fig:lamellar_perp_kinetics} we display the resulting temporal behavior for Pe = 0.25 in the $x$-$z$-plane (the configuration is transitionally invariant along the $y$-direction). The color-code is the same as in Fig. \ref{Fig:lamellar_highPe}, and the plotted times are $t^* = 0$ (a), 0.5 (b), 1.0 (c), 2.0 (d), 4.0 (e), and 8.0 (f). For short times $t^* \lesssim$ 1.0 we see that the lamellae are increasingly distorted by the flow, but they are not completely destroyed (Figs. \ref{Fig:lamellar_perp_kinetics} (a)--(c)). At some point in time, however, there is a kinetic phase-transition where the initial structure is completely destroyed. Close to the wall boundary particles start to form stripes, and in between $\rho(\mathbf{r})$ shows a \lq turbulent\rq\, behavior (Fig. \ref{Fig:lamellar_perp_kinetics} (d)). Subsequently, the particles self-aggregate into stripes parallel to the flow direction, which finally results in a perfect parallel-oriented L structure (see Fig. \ref{Fig:lamellar_perp_kinetics} (e)--(f)). In \cite{SupplementalMaterial1} we show a movie to further illustrate the behavior of the transversal L phase under shear up to times $t^* = 8.0$. When shear is switched off, unlike to the case of high shear, the system stays in the configuration of parallel lamellae. This is indeed not too surprising as for the chosen state point the parallel configuration is the most stable configuration (as discussed at the beginning of Sec. \ref{SubSubSec_parallel_L}).

\subsection{Gyroid phase under steady shear} \label{SubSec_gyroid}
\begin{figure}[t!] 
	\centering
	\includegraphics[width = 8.5cm]{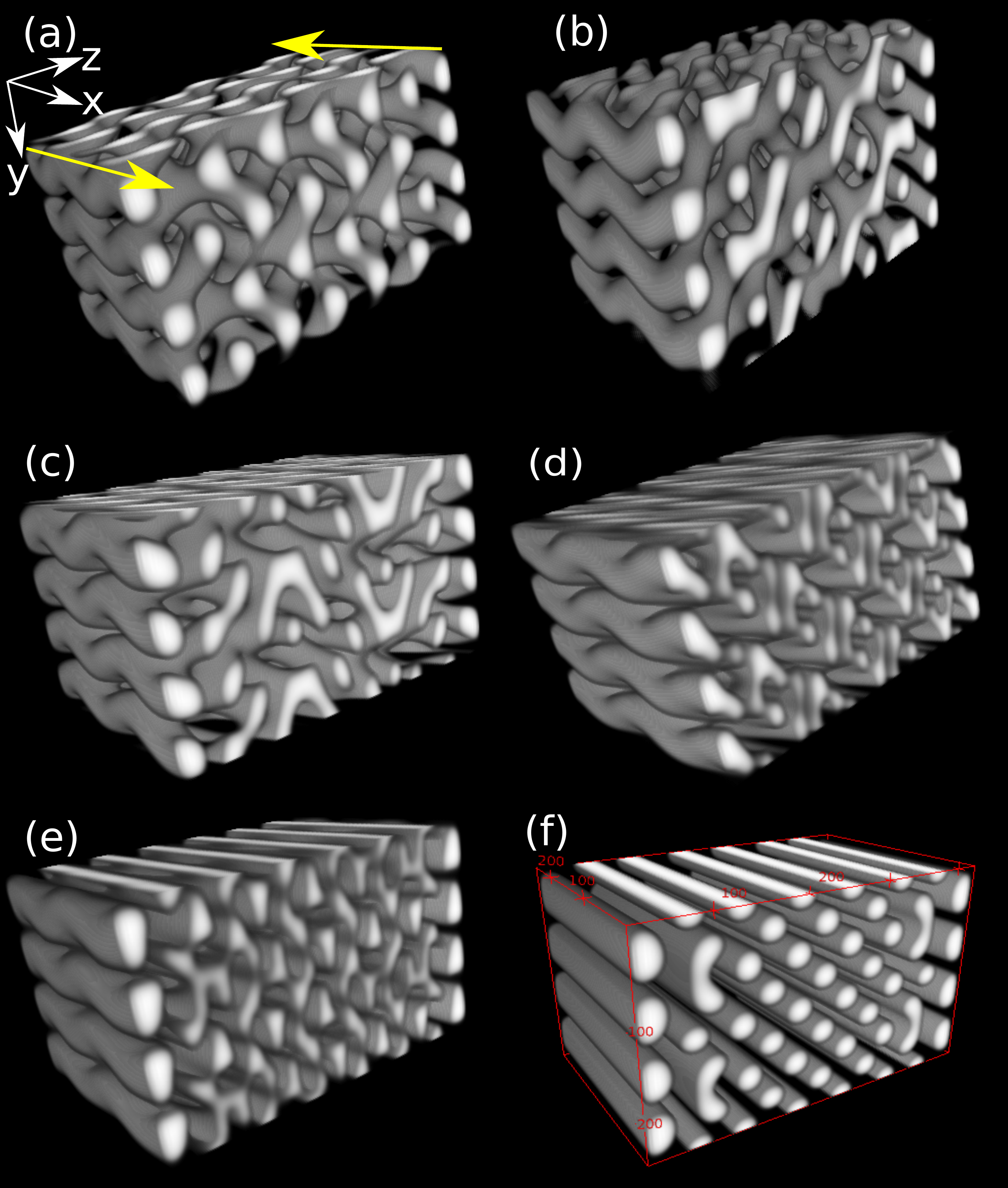} 
	\caption{Gyroid structure between repulsive walls under steady shear with $\text{Pe}=0.25$ at times: $t^* = 0.0$ (a),  $0.5$ (b), $1.0$ (c), $2.0$ (d), $4.0$ (e), and 12.0 (f). Yellow arrows indicate shear flow along $x$-direction. }
	\label{Fig:gyroid_time}
\end{figure}
\begin{figure}[t!] 
	\centering
	\vspace{-0.5cm}
	\includegraphics[width = 8.5cm]{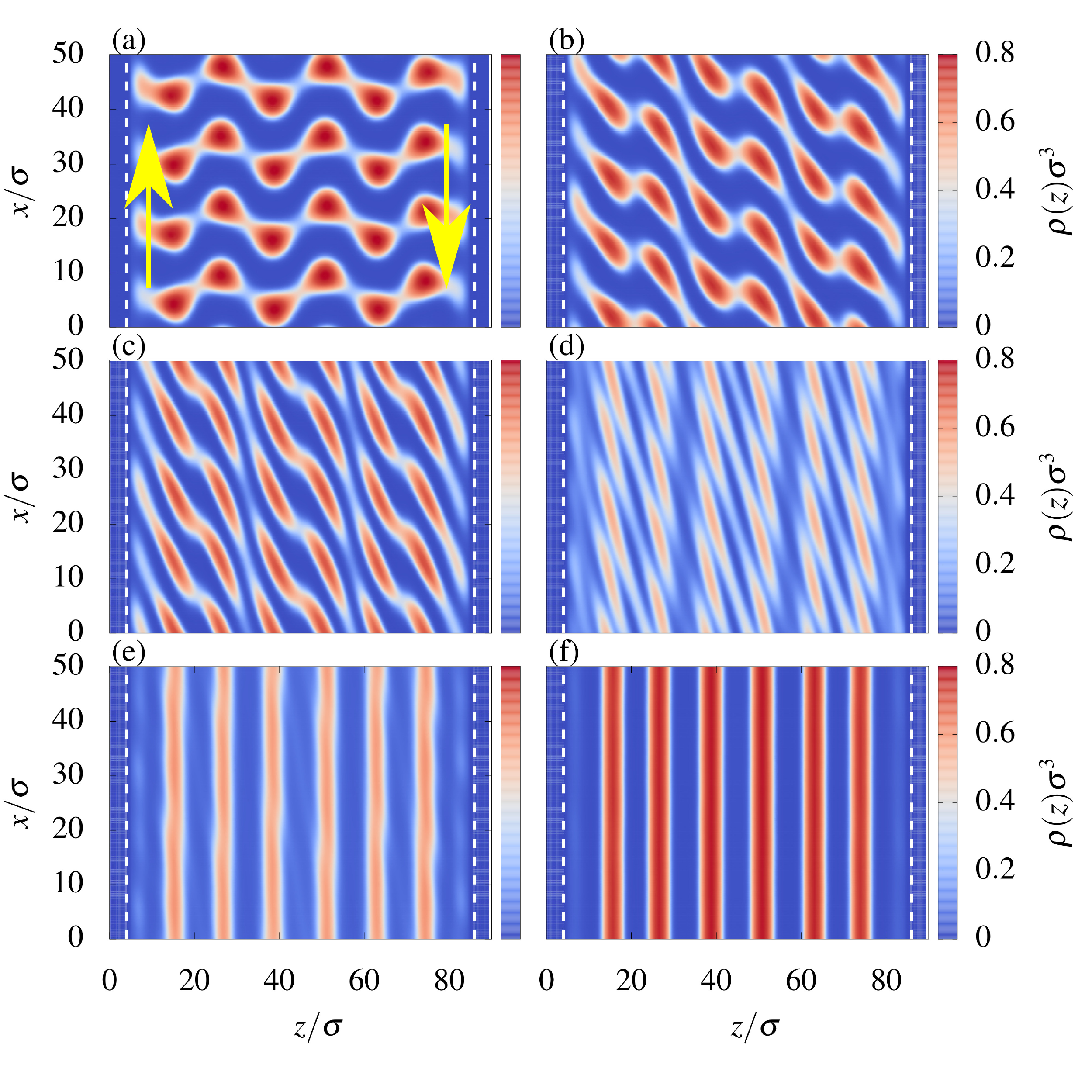} 
	\caption{$x$-$z$-plane view of the sheared gyroid at a constant $y$-coordinate for times 0.0 (a), 0.5 (b), 1.0 (c), 2.0 (d), 4.0 (e), and 12.0 (f). The dashed lines are guides to the eye marking the position of the repulsive walls, yellow arrows in (a) indicate the shear flow, and in \cite{SupplementalMaterial1} a movie illustration can be found.}
	\label{Fig:gyroid_2d}
\end{figure}

In this subsection we examine the behavior of the G phase under shear, a structure that obeys a significantly more complex morphology than a lamellar, 2D-hexagonal or BCC phase. We consider two cases: first, a moderate shear with Pe = 0.25 as in the previous section, and a very weak shear with Pe = 0.01.
In Fig. \ref{Fig:gyroid_time} we show the time evolution of the full 3D density under shear with Pe = 0.25 for times $t^* = 0.0$ (a), 0.5 (b), 1.0 (c), 2.0 (d), 4.0 (e) and 12.0 (f), and in Fig. \ref{Fig:gyroid_2d} the corresponding $x$-$z$-plane view is displayed for the same times. The color code is as in Fig. \ref{Fig:lamellar_highPe}, and as for the transversal L phase, in \cite{SupplementalMaterial1} a movie is provided up to times $t^* = 14.0$. As for the transversal-oriented L phase, the gyroid becomes increasingly distorted along the flow direction, where up to $t^* \lesssim 1.0$ the morphology with a highly interconnected network is clearly identifiable. However, in the $y$-$z$-plane, normal to the flow direction, stronger deformations are emerging. For longer times, the gyroid morphology is no longer visible, and is replaced by a turbulent-like system of small tubes and lamellae where interconnections perpendicular to the flow direction nearly vanished. Subsequently, this structure reorders into 2D-hexagonally arranged cylinders, with translational invariance along the flow direction (Figs. \ref{Fig:gyroid_time} (d)--(f)), although some deviations are present close to the barrier boundaries. When switching shear off after reaching the final state, we find that the system is trapped in a metastable thermodynamic minimum: the density distribution does not change anymore when time goes by. This can be attributed to a rather complex free-energy landscape -- while for the chosen values of $\eta, A, B, z_1$ and $z_2$ (see Sec. \ref{SubSec_GyroidWall}) the initial G represents the \textit{most} stable microphase (cf. Fig. \ref{Fig_phaseDiag}), the HEX phase can also exist but is metastable w.r.t. the G phase.  Furthermore, one should bear in mind that a general property of DDFT is that random-noise fluctuations are treated in a mean-field fashion, since fluctuations are averaged out during the derivation of DDFT from the underlying microscopic stochastic equations \cite{MarconiTarazona1999, ArcherEvans2004}. As such, the system cannot overcome possible energy barriers between distinct types of cluster states by simply \lq waiting\rq\, for a sufficiently long time period in order to minimize its free energy; moreover it is a priori not clear whether stochastic fluctuations would be sufficient to drive the system towards its global energetic minimum. 

Hence, applying shear drives a topological phase transition G $\rightarrow$ HEX, thereby dramatically changing the Euler-characteristic $\chi_\text{G}$ of the initial structure; $\chi_\text{G}$ is strongly negative due to the highly inter-connective network, while the final structure presumably obeys a very small Euler characteristic $\chi_\text{hex}$, as the latter would be zero for perfect cylinders without any defects. 
What is striking about the present results is that similar shear-induced instabilities causing topological phase transitions G $\rightarrow$ HEX have been observed for block copolymers in both computer simulations \cite{Pinna2008} and experiments \cite{Eskimergen2005}. Moreover, transitions from transversal L $\rightarrow$ parallel L structures under weak shear have also been reported \cite{Peters2012}. While it is well known that equilibrium bulk properties of self-organized cluster phases of colloidal systems with competing interactions and block copolymers obey a universal behavior, for out-of-equilibrium phenomena this has not been reported before. Considering the simulation study of sheared block copolymers in Ref. \cite{Pinna2008}, even the kinetic pathway (for interested readers cf. Figs. 6,7, and 8 in Ref.  \cite{Pinna2008}) reveals intriguing analogies with our findings. 
These parallels can be explained by rationalizing that dynamic processes in block copolymer systems often make use of equations of motion that (from a mathematical point of view) are similar to the colloidal DDFT equations (see, e.g., Ref. \cite{Huinink2000}; further references are given in Ref. \cite{Register2013}).
\begin{figure}[t!] 
	\centering
	\includegraphics[width = 8.5cm]{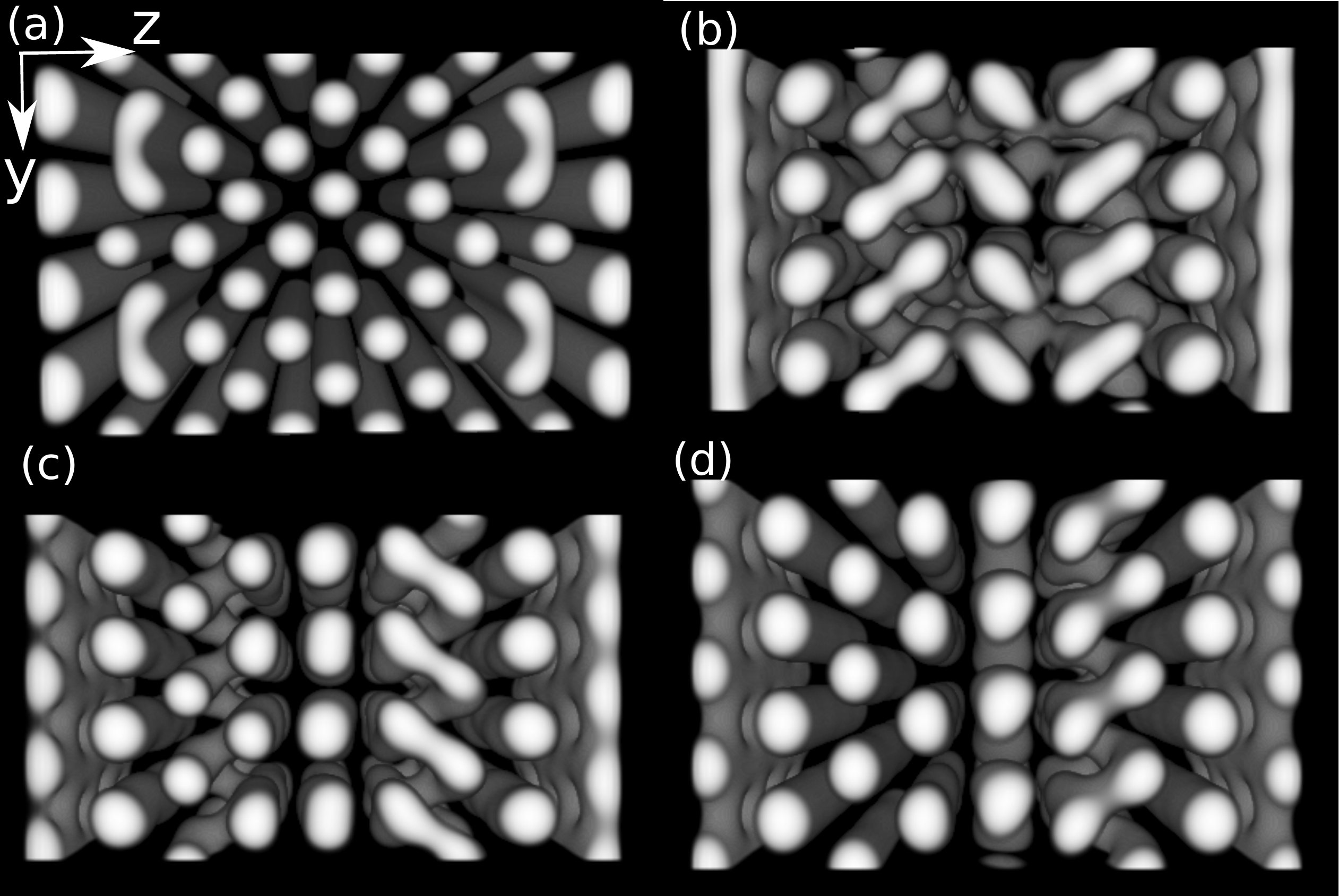} 
	\caption{(a) 3D-frontal view on the $y$-$z$-plane at a constant $x$-coordinate showing the steady-state configuration of the sheared gyroid phase with Pe = 0.25. (b)-(d) Same as in (a) but with Pe = 0.01 for times $t^* = 200$ (b), 300 (c) and 400 (d).}
	\label{Fig:gyroid_low_high_shear}
\end{figure}
In Fig. \ref{Fig:gyroid_low_high_shear} we compare the final state (reached at $t^* \approx 12$) of the sheared G phase for Pe = 0.25 (a) to long-time results at $t^* = 200$ (b), 300 (c), and 400 (d) for Pe = 0.01. Calculation of these results took more than 1 month of computation time. Note that the total shear strain $\text{Pe}\cdot t$ for (a) and (c) is the same, yet the impact on density is significantly different.  For the latter, we see that (for all times) the shear is too weak to destroy interconnections between the cylinders located perpendicular to the flow direction. 
These results beg the question of whether a minimal shear rate exists, below which the system cannot be driven out of its (local) thermodynamic minimum, i.e., below which the initial Euler characteristic does not change significantly over time. We think that such a minimal shear rate exists, but presumably it can become very small with a corresponding Peclet number of order $10^{-3}$ -- $10^{-5}$ or even below. 
Unfortunately, this cannot be validated straight forwardly, as for such small shear rates possible changes of the density due to shear in turn will take place on time scales of $10^3 \tau_B$ -- $10^5\tau_B$ which would require several months or years of computation time. Thus, at this point we cannot provide an proper answer to this question. One may also address this question by analytically expanding the DDFT in the limit of very low shear rates around an equilibrium state such as the G phase, which, however, is beyond the scope of the current work.

\section{Summary and Outlook} \label{SecDiscussion}

In this work, we employed classical density functional theory and its dynamical extension to investigate the effect of (steady) shear flow on self-assembled colloidal microphases. The latter are periodically ordered cluster phases that arise due to a competition between attractive and repulsive interparticle forces; the loss in entropy due to spatially ordering is (over-)compensated by a gain in configurational energy. Remarkably, such clusters can show complex non-spherical morphologies such as Gyroid phases, albeit the underlying particle interactions can be spherically-symmetric \cite{EdelmannRothPRE2016, Charbonneau2016_PRL}. In particular, the bulk phase behavior of the present system shows intriguing similarities with block copolymers \cite{Foerster1994_copolymerClusterPhases, MatsenSchick1994_PRL, Ciach2008, Ciach2013}, and
here we demonstrated that these similarities are not bound to equilibrium situations: upon applying a steady shear flow, the system is driven out of equilibrium which results in kinetic phase-transitions depending on the respective Peclet number. For sufficiently weak shear with Pe < 1, we showed that one can induce e.g. a G $\rightarrow$ HEX transition, which both in experiments and computer simulations \cite{Eskimergen2005, Pinna2008, Register2013} has been observed for block copolymers. Similarly, a transversal oriented L phase  (lamellae perpendicular to wall boundaries and flow direction) reorientates itself under shear into a parallel L structure, which also was observed in particle dynamics simulations of block copolymers \cite{Peters2012}. 
In contrast, at higher shear rates with Pe > 1, we found that clusters are destroyed and the fluid obeys a phase transition into an unordered, dissolved phase; this is a direct consequence of the hard-core interactions between particles, and in this work is accounted for by a recent version of dynamical DFT \cite{ScacchiKruegerBrader2016}, adequately capturing effects of sheared colloidal suspensions. In particular, in this regime the behavior seems to be different to block copolymers: upon increasing shear, the latter typically do not dissolve but rather tend to form distinct clusters compared to lower shear rates \cite{Peters2012}. The present results open perspectives to future work, where the regime Pe > 1 may be studied in more detail. Here, it would be interesting to study the behavior of more complex structures such as a gyroid or BCC phase under high shear; another interesting (and important) point is to go beyond the flow kernel for pure hard spheres. However, these points require that very high computer resources are available. In the high-shear regime, microphases need to be described on three-dimensional grids with spatial resolutions that allow to properly account for packing effects on the one-particle scale (see Secs. \ref{SubSec_GyroidWall} and \ref{SubSec_Res_LamellarShear}). However, up to this date, the treatment of very large systems with domain sizes consisting of $\sim (1024)^3$ grid points seems to be out of reach within the framework of (dynamic) DFT, even when employing modern high-performance graphics cards to significantly speedup calculations as we have done in this work.

Finally, it is important to note that in this work we neglected hydrodynamic interactions (HI) i.e., we assumed that the colloidal particles instantaneously follow the flow field without distorting the latter significantly. A coupling between particles and solvent may be incorporated via solving the 3D Stokes equations for incompressible flow
\begin{equation} \label{Eq:StokesEquation}
	\eta \nabla^2 \mathbf{v}(\mathbf{r}) - \nabla p = -\rho(\mathbf{r})\mathbf{f}(\mathbf{r})\,,
\end{equation}
where $\eta$ is dynamic viscosity, $p$ a pressure field enforcing the incompressibility constraint, and $\mathbf{f}(\mathbf{r})$ is a body force exerted on the solvent, in the present context given by 
\begin{equation} \label{Eq:BodyForce}
	\mathbf{f}(\mathbf{r}) = -\nabla \frac{\beta\Omega[\rho]}{\delta\rho(\mathbf{r})}\,.
\end{equation}
Equations \eqref{Eq:StokesEquation} and \eqref{Eq:BodyForce} may then be solved self-consistently with suitable boundary conditions along with Eq. \eqref{EqDDFT_flowfield}. Preliminary results for Pe < 1 (here we completely neglected the flow kernel as incorporating both of the latter and HI is numerically extremely expensive) indicate that HI seem to not impact significantly the phase transitions shown in this work. For higher shear rates (e.g. Pe = 1.5 while neglecting the flow kernel), we obtain a non-linear, slightly oscillatory flow profile along the shear-gradient axis similar to results provided in Ref. \cite{Peters2012} for block copolymers. However, a more detailed analysis of the influence of HI, in particular at higher shear rates, will be subject of future work.

\section*{Acknowledgements}
Funding by the Carl-Zeiss-Stiftung is gratefully acknowledged. We thank M. Kr\"uger for stimulating discussions. We further thank S.C. Lin and M. Oettel for discussions on the exponential time difference algorithm. 

\bibliographystyle{apsrev}
\bibliography{references}

\appendix

\section{Exponential Time Differencing} \label{Appendix_A}
For simplicity, we  consider the standard DDFT equation without any external flow (that can be included straight forwardly) for a single-component system in three dimensions, which is given by \cite{MarconiTarazona1999, ArcherEvans2004}
\begin{align} \label{EqDDFT_Appendix}
	\frac{1}{D_0}\frac{\partial \rho}{\partial t} &= \nabla^2 \rho(\mathbf{r},t) - \nabla\left(\rho(\mathbf{r},t)\nabla c^{(1)}(\mathbf{r},t)\right) \notag \\ &+ \nabla\left(\rho(\mathbf{r},t)\nabla \beta V_\text{ext}(\mathbf{r},t)\right)\,.
\end{align}
We present a numerical time integrator employed in this work that turned out to be very accurate for diffusive-type partial differential equations such as Eq. \eqref{EqDDFT_Appendix}. The method, known as exponential time differencing \cite{CoxMatthews2002}, is a numerical integrator that treats the diffusive term $\sim \nabla^2 \rho$ \textit{exactly}. 
 In Fourier-space, the DDFT equation can be written as
\begin{equation} \label{Eq_DDFT_Fourier}
	\frac{\partial \rho_k}{\partial t} = \mathcal{L}_k \rho_k + \mathcal{N}_k\,,
\end{equation}
where $\mathcal{L}_k := -D_0|\mathbf{k}|^2$ is the Laplace operator in Fourier-space, and $\mathcal{N}_k$ contains the Fourier-transform of the remaining terms (typically non-linear in the density) of Eq. \eqref{EqDDFT_Appendix}. For sake of readability in this section we write $f_k \equiv \widehat{f}(\mathbf{k})$, where $\widehat{f}(\mathbf{k})$ denotes the three-dimensional Fourier-transform of a function $f(\mathbf{r})$. Thus, $k$ denotes \textit{not} the absolute value of $\mathbf{k}$, but an index describing a specific Fourier-mode $\mathbf{k} = (k_x, k_y, k_z)$. 

Equation \eqref{Eq_DDFT_Fourier} can be rearranged as follows
\begin{equation}
	\frac{\partial}{\partial t}\left( \rho_k e^{-\mathcal{L}_k t}\right) = \left(\frac{\partial \rho_k}{\partial t} - \mathcal{L}_k \rho_k\right)e^{-\mathcal{L}_k t} = \mathcal{N}_k e^{-\mathcal{L}_k t}\,.
\end{equation}

Integrating over a small time step $\Delta t$, we obtain the following  relation
\begin{align}
	\int_{t}^{t+\Delta t} \text{d}t'\,\frac{\partial}{\partial t'} \left(\rho_k e^{-\mathcal{L}_k t'}\right) = \int_{t}^{t + \Delta t} \text{d}t'\, \mathcal{N}_k(t') e^{-\mathcal{L}_k t'}\,,
\end{align}
which yields the \textit{exact} result
\begin{equation}
	\rho_k(t + \Delta t) = \rho_k(t) e^{\mathcal{L}_k \Delta t} + e^{\mathcal{L}_k (t+\Delta t)}\int_{t}^{t + \Delta t} \text{d}t'\, \mathcal{N}_k(t')e^{-\mathcal{L}_k t'}\,.
\end{equation}
One can now approximate the integrand 
\begin{align}
	\mathcal{N}_k(t') &= \mathcal{N}_k(t) + \frac{\mathcal{N}_k(t) - \mathcal{N}_k(t - \Delta t)}{\Delta t}(t' - t)\notag\\ &+ \mathcal{O}((t' - t)^2)\,,
\end{align}
giving rise to
\begin{equation}
	\rho_k(t + \Delta t) =  \rho_k(t) e^{\mathcal{L}_k \Delta t} + I_1 + I_2\,,
\end{equation}
where
\begin{align}
	I_1 &\equiv \frac{\mathcal{N}_k}{\mathcal{L}_k}\left(e^{\mathcal{L}_k \Delta t} - 1\right)\notag\,, \\
	I_2 &\equiv \left(\frac{\mathcal{N}_k(t) - \mathcal{N}_k(t - \Delta t)}{\Delta t}\right)\left(\frac{-\Delta t}{\mathcal{L}_k} - \frac{1}{\mathcal{L}_k^2}\left(1 - e^{\mathcal{L}_k \Delta t}\right)\right)\,.
\end{align}
Neglecting $I_2$ in most cases provides already a very accurate approximation to the dynamics described with DDFT, and is used throughout this work. In particular, the implementation effort is comparable to that of a simple Euler-forward algorithm. We found that this time integrator stays stable (and accurate, checked by iterating equilibrium density profiles) for very long times $t \sim 10^3 \tau_B$ using time steps of $\Delta t = 10^{-4}$ in three dimensions, and up to $t \sim 10^5 \tau_B$ employing $\Delta t = 10^{-3} \tau_B$ in (effectively) one dimensional situations (tested for sheared hard-spheres between planar hard walls) and a grid spacing of $\delta = R/10$. This numerical integrator may also be important for situations where local chemical potential gradients become very small resulting in slow particle dynamics.

\section{Fourier-transform of the flow kernel $\mathbf{K}(\mathbf{r})$ for hard spheres} \label{Appendix_B}
For shear flow which flows in $x$-direction and has a gradient along the $z$-axis, the flow kernel for hard spheres can be written as \cite{ScacchiKruegerBrader2016}
\begin{align}
	\mathbf{K}(\mathbf{r}) &= -D_0 \left(\text{Pe}\frac{xz}{r^2}h_1(r)\right. \\ &+ \text{Pe}^2\left(\frac{x^2z^2}{r^4}h_2(r) + \frac{x^2-z^2}{4r^2}h_3(r)\right. \notag \\ &+ 
	\left.\left.\frac{x^2+z^2}{4r^2}h_4(r) + \frac{1}{2}h_5(r)\right)\right)\nabla \exp(-\beta u_\text{hs}(r)) \\
	&\equiv \mathbf{K}_1(\mathbf{r}) + \mathbf{K}_2(\mathbf{r}) + \mathbf{K}_3(\mathbf{r}) + \mathbf{K}_4(\mathbf{r}) + \mathbf{K}_5(\mathbf{r})\notag \,,
\end{align}
where the functions $h_i(r)$ are given by
\begin{align}
	h_1(r) &= \frac{16}{3\tilde{r}^3}\,,\\
	h_2(r) &= \frac{2}{3}\left(\frac{1}{\tilde{r}} - \frac{16}{\tilde{r}^5}\right)\,,  \\
	h_3(r) &= \frac{8}{27}\left(\frac{3}{\tilde{r}} - \frac{4}{\tilde{r}^3} \right) \,,
\end{align}
and
\begin{align}
		h_4(r) &= -\frac{32}{63}\left(\frac{5}{\tilde{r}^3} - \frac{12}{\tilde{r}^5}\right)\,,\\ 
		h_5(r) &= -\frac{4}{945}\left(\frac{105}{\tilde{r}} - \frac{200}{\tilde{r}^3} + \frac{144}{\tilde{r}^5}\right)\,,
\end{align}
and $\tilde{r} = r/R$. Using the flow kernel for hard spheres as a \lq zero-order\rq\, perturbation for the present system, we can calculate analytically the Fourier-transform of $\mathbf{K}(\mathbf{r})$ which is needed for an efficient calculation of the flow field $\mathbf{v}_\text{fl}(\mathbf{r})$. To this end, note that $\nabla \exp(-\beta u_\text{hs}(r)) = \delta(r - \sigma)\mathbf{r}/r$ and
\begin{equation}
	x_\alpha^n \exp(-i\mathbf{k}\cdot\mathbf{r}) = i^n\frac{\partial ^n}{\partial k_\alpha^n} \exp(-i\mathbf{k}\cdot\mathbf{r})\,,
\end{equation}
where $n\in\mathbb{N}$ denotes an arbitrary exponent and $x_\alpha=x_1,x_2,x_3$ correspond to the spatial directions $x,y,z$. Hence, the Fourier-transform of e.g. the $\alpha$-component of $\mathbf{K}_1(\mathbf{r})$ is given by $(k=|\mathbf{k}|)$
\begin{align}
	\widehat{\mathbf{K}}_1^\alpha(\mathbf{k})& = -D_0 \text{Pe} \int \text{d}\mathbf{r}\,\exp(-i\mathbf{k}\cdot\mathbf{r}) \frac{x_1x_3}{r^2} h_1(r) \delta(r-\sigma) \frac{x_\alpha}{r}\notag\\
	&= i D_0\text{Pe}\frac{\partial^3}{\partial k_{1}\partial k_{3}\partial k_{\alpha}} \left( \frac{4\pi}{k}\int\text{d}r\sin(kr)\frac{h_1(r)}{r^2}\delta(r-\sigma) \right)\notag\\
	&= i D_0\text{Pe}\frac{\partial^3}{\partial k_{1}\partial k_{3}\partial k_{\alpha}}\left(\frac{4\pi}{k}\sin(\sigma k)\frac{h_1(\sigma)}{\sigma^2}\right)\,,
\end{align}
which can be evaluated analytically. Similar, we obtain
\begin{align}
	\frac{\widehat{\mathbf{K}}_2^\alpha(\mathbf{k})}{D_0 \text{Pe}^2} &= -i\frac{\partial^2}{\partial k_1^2}\frac{\partial^2}{\partial k_3^2} \frac{\partial}{\partial k_\alpha}\left(\frac{4\pi}{k} \frac{\sin(\sigma k)h_2(\sigma)}{\sigma^4}\right)\,,\\
		\frac{\widehat{\mathbf{K}}_3^\alpha(\mathbf{k})}{D_0 \text{Pe}^2} &= i\left(\frac{\partial^2}{\partial k_1^2} - \frac{\partial^2}{\partial k_3^2}\right)\frac{\partial}{\partial k_\alpha}\left(\frac{4\pi}{k} \frac{\sin(\sigma k)h_3(\sigma)}{4\sigma^2}\right)\,,\\
		\frac{\widehat{\mathbf{K}}_4^\alpha(\mathbf{k})}{D_0 \text{Pe}^2} &= i\left(\frac{\partial^2}{\partial k_1^2} + \frac{\partial^2}{\partial k_3^2}\right)\frac{\partial}{\partial k_\alpha}\left(\frac{4\pi}{k} \frac{\sin(\sigma k)h_4(\sigma)}{4\sigma^2}\right)\,,\\
		\frac{\widehat{\mathbf{K}}_5^\alpha(\mathbf{k})}{D_0 \text{Pe}^2} &= -i\frac{\partial}{\partial k_\alpha}\frac{1}{2}\left(\frac{4\pi}{k} \sin(\sigma k)h_5(\sigma)\right)\,.
\end{align}

\end{document}